\documentclass[%
 aip,
rsi,
 amsmath,amssymb,
 reprint,%
 colorlinks=true,
 linkcolor=blue,
 urlcolor=blue,
 citecolor=blue
]{revtex4-1}

\usepackage{mathrsfs}
\usepackage{hyperref}

\usepackage{graphicx}
\usepackage{dcolumn}
\usepackage{bm}
\usepackage{url}
\usepackage{amssymb}
\usepackage{amsmath}
\usepackage{mathtools}
\usepackage{braket}
\usepackage[mathlines]{lineno}
\usepackage{xcolor}

\usepackage[utf8]{inputenc}
\usepackage[T1]{fontenc}
\usepackage{mathptmx}
\usepackage{etoolbox}

\newcommand{\unit}[1]{\ \mathrm{#1}}

\makeatletter
\def\@email#1#2{%
 \endgroup
 \patchcmd{\titleblock@produce}
  {\frontmatter@RRAPformat}
  {\frontmatter@RRAPformat{\produce@RRAP{*#1\href{mailto:#2}{#2}}}\frontmatter@RRAPformat}
  {}{}
}%
\makeatother
\begin{document}

\preprint{AIP/123-QED}

\title[]{Development of a polarimetry method toward \textit{in-situ} substrate birefringence characterization of ground-based gravitational wave detectors}
\author{Satoshi Tanioka}
\affiliation{School of Physics and Astronomy, Cardiff University, Cardiff CF24 3AA, United Kingdom}

\email{TaniokaS@cardiff.ac.uk}
\author{Terri Pearce}

\author{Keiko Kokeyama}
\affiliation{School of Physics and Astronomy, Cardiff University, Cardiff CF24 3AA, United Kingdom}

\date{\today}

\begin{abstract}

Improving the sensitivity of gravitational wave detectors is necessary to enrich scientific outcome of gravitational wave astronomy.
Birefringence in test mass mirrors of gravitational wave detectors can become an important factor for both current and next-generation gravitational wave detectors to achieve improved performance.
In-situ birefringence characterization can become an essential diagnostic tool for detector performance, and needs to be established.
We report a possible in-situ birefringence characterization method and its experimental results with a tabletop setup.
The scheme proposed and demonstrated in this paper can be used as a diagnostic tool in large-scale gravitational wave detectors.
We also discuss possible technological developments toward implementation in future gravitational wave detectors.

\end{abstract}

\maketitle

\section{
\label{sec:intro}
Introduction}

Ground-based laser interferometric gravitational wave detectors (GWDs) have opened a new window on the Universe by enabling direct detection of gravitational waves (GWs). \cite{Abbott2016, Abbott2017}
Those GWDs consist of dual-recycled Fabry-Perot Michelson interferometers to achieve sufficient sensitivity to enable direct detection of GWs. \cite{Aso2013, Aasi2015, Acernese2015}
In order to enhance the GW detection rate and their scientific outcome, various research activities are being performed both on-site and off-site. \cite{Soni2021, Vajente2021, Capote2025}
Furthermore, next-generation GWDs, such as Cosmic Explorer and the Einstein Telescope, are expected to largely enhance the scientific impact of gravitational wave astronomy by having better sensitivities. \cite{Punturo2010, Hall2021}

Development of high-quality test mass mirrors will play a crucial role in both improving the performance of current GWDs and in the realization of the next-generation GWDs. \cite{Degallaix2019}
One of the important properties of such test mass mirrors is birefringence distribution.
Recent studies in KAGRA have revealed that inhomogeneous birefringence in sapphire input test masses (ITMs) degrade the performance of the detector. \cite{Somiya2019, Hirose2020}
Their typical differential phase retardation induced by birefringence were order of $100\unit{nm}$ per single trip, which corresponds to about $0.6\unit{rad}$. \cite{Wang2024}
Such birefringence in an ITM is known to affect the detector performance as it can degrade the power-recycling gain (PRG), resulting in reduced circulating power. \cite{Winkler1994}
In addition to the PRG reduction, the ITM birefringence can affect the contrast of the interferometer, and reduce the achievable squeezing level. \cite{Aasi2015, Krüger2016, Michimura2024}

While a tabletop setup can characterize the test mass birefringence, the actual birefringence distribution can be altered by a stress induced by a suspension or by cooling down to cryogenic temperature.
Specifically, next-generation GWDs will employ a few hundred kilogram test masses which may be susceptible to substantial stress, resulting in stress-induced birefringence. \cite{Punturo2010, Hall2021}
This might not satisfy the requirement to achieve the designed performance of GWDs.
In addition, their test masses are thicker than the current test masses which can therefore enhance the phase retardation as shown in Eq. (\ref{eq:retardation}).
Therefore, in-situ birefringence characterization will play an important role in understanding the properties of test masses and evaluate the performance of the GWDs.

The test mass birefringence can be characterized by rotating either the input beam polarization or the sample mirror itself in tabletop experiments. \cite{Wang1999, Zeidler2023, Wang2024, Singh2025}
However, neither can be rotated in GWDs.
Therefore, in order to realize the in-situ birefringence characterization, it is necessary to develop a method which does not require rotation of the input beam polarization or the test mass.

In this paper, we propose and demonstrate a polarimetry approach to enable in-situ test mass birefringence characterization in a gravitational wave detector.
We have developed a tabletop setup which shows reasonable repeatability and measurement results.
We also discuss further developments toward implementing the proposed birefringence characterization scheme to GWDs.

\section{Birefringence calculation}


\subsection{Definition}

\begin{figure}[htbp]
\includegraphics[width=8.6cm]{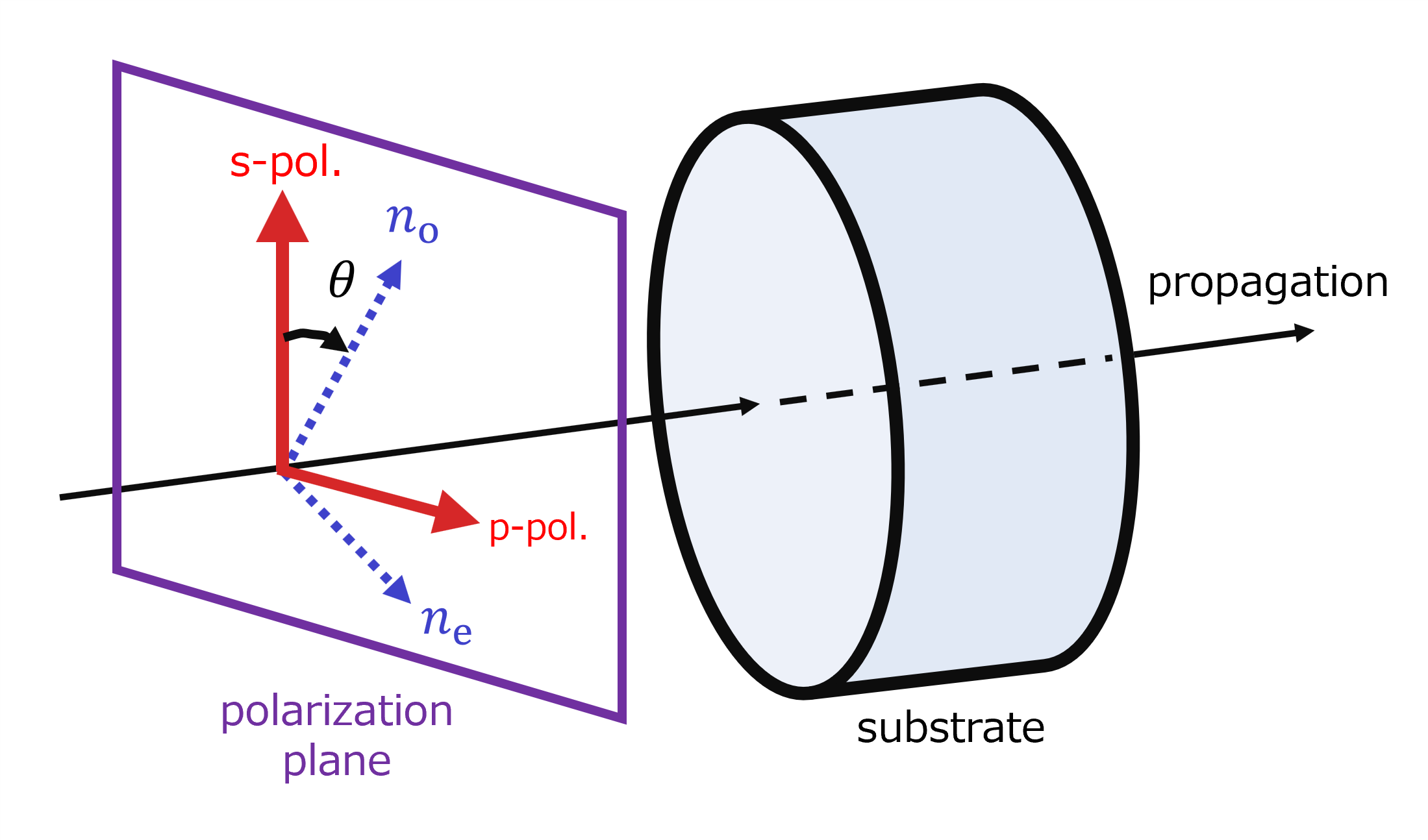}
\caption{\label{fig:definition} 
Schematic of the definition.
The mirror substrate has two orthogonal refractive index axes, $n_{\mathrm{o}}$ and $n_{\mathrm{e}}$, which are rotated by $\theta$ from s- and p-polarization axes, respectively.
}
\end{figure}

We define parameters and geometry to characterize the birefringence distribution as shown in Fig. \ref{fig:definition}.
In the same manner as the previous work, we set s- and p-polarization axes in the vertical and horizontal directions with respect to the laboratory frame. \cite{Wang2024}
The polarization state of the light field can be expressed as
\begin{align}
    \ket{E} =
    \begin{pmatrix}
    E_{\mathrm{s}} \\
    E_{\mathrm{p}}\\
    \end{pmatrix},
\end{align}
where $E_{\mathrm{s}}$ and $E_{\mathrm{p}}$ are the electric field of the s- and p-polarized fields, respectively.

When a substrate material exhibits birefringence, that property can be often described by refractive index difference, $\Delta n$, between two distinctive orthogonal axes, $n_{\mathrm{o}}$ and $n_{\mathrm{e}}$.
These two orthogonal refractive index axes are respectively called ordinary and extraordinary axes.
Here we assume that these axes are rotated by an angle of $\theta$ with respect to the polarization axes as shown in Fig. \ref{fig:definition}.

The polarization state of the light field can be modified when it passes through the birefringent substrate.
Similar to the previous work, we define the following common and differential phase changes as \cite{Wang2024}
\begin{align}
    \alpha_+ \coloneqq  \frac{\pi d}{\lambda}\left(n_{\mathrm{o}}+n_{\mathrm{e}}\right), \\
    \alpha_- \coloneqq \frac{\pi d}{\lambda}\left(n_{\mathrm{o}}-n_{\mathrm{e}}\right), \label{eq:retardation}
\end{align}
where $d$ is the thickness of the sample and $\lambda$ is the laser wavelength.
By using these definitions, the Jones matrix of the birefringent sample, $\hat{M}$ can be expressed as
\begin{align}
    \hat{M}
    &= \mathrm{e}^{i\alpha_+}
    \begin{pmatrix}
        \cos\alpha_- + i\sin\alpha_-\cos2\theta & i\sin\alpha_-\sin2\theta \\
        i\sin\alpha_-\sin2\theta & \cos\alpha_- - i\sin\alpha_-\cos2\theta \\
    \end{pmatrix}.
    \label{eq:mirror}
\end{align}
In this convention, the range of the orientation and differential phase retardation are $-45\unit{deg}\leq\theta\leq45\unit{deg}$ and $-\pi/2\leq\alpha_-\leq\pi/2$, respectively.
Note that the sign of $i\sin\alpha_-\sin2\theta$ is flipped compared to the previous work due to the difference in the definition of $\theta$. \cite{Wang2024}


In our setup, the s-polarized beam is used as an input beam in the same manner as in KAGRA. \cite{Akutsu2021}
The electric field of the input beam can be described as
\begin{align}
    \ket{E_{\mathrm{in}}} &=
    \begin{pmatrix}
    E_{\mathrm{in}} \\
    0 \\
    \end{pmatrix},
\end{align}
where $E_{\mathrm{in}}\coloneqq  E_0\mathrm{e}^{i\Omega t}$, $E_0\in \mathbb{C}$ is the amplitude of the field, and $\Omega$ is the laser angular frequency.
When the s-polarized beam passes the birefringent sample, the transmitted light field becomes
\begin{align}
    \hat{M}\ket{E_{\mathrm{in}}} &= E_{\mathrm{in}}\mathrm{e}^{i\alpha_+}
    \begin{pmatrix}
        \cos\alpha_- + i\sin\alpha_-\cos2\theta \\
        i\sin\alpha_-\sin2\theta \\
    \end{pmatrix}.
    \label{eq:aftersample}
\end{align}
Therefore, the input beam field is scattered into orthogonal polarization depending on the parameters of the birefringent sample $\alpha_-$ and $\theta$.

\subsection{Characterization method}

In this subsection, we describe the birefringence characterization method.
Instead of rotating the input beam polarization or the birefringent sample, we rotate the polarization of the transmitted beam through the sample.

In our setup , as shown in Fig. \ref{fig:PPC}, a half-wave plate (HWP) is inserted right after the sample.
The Jones matrix of a HWP can be expressed as
\begin{align}
    \hat{M}_{\mathrm{HWP}}(\theta_{\mathrm{H}}) &=
    \begin{pmatrix}
        \cos2\theta_{\mathrm{H}} & \sin2\theta_{\mathrm{H}} \\
        \sin2\theta_{\mathrm{H}} & -\cos2\theta_{\mathrm{H}} \\
    \end{pmatrix},
\end{align}
where $\theta_{\mathrm{H}}$ is the orientation of the HWP.
Then, the field after the HWP becomes
\begin{align}
    \ket{E_{\mathrm{out}}(\theta_{\mathrm{H}})} &= \hat{M}_{\mathrm{HWP}}(\theta_{\mathrm{H}})\hat{M}\ket{E_{\mathrm{in}}}.
    \label{eq:output}
\end{align}

First, we consider the case when the HWP orientation is aligned to the horizontal and vertical directions, i.e., $\theta_{\mathrm{H}}=0$.
The Jones matrix of the HWP becomes
\begin{align}
    \hat{M}_{\mathrm{HWP}}(0) &=
    \begin{pmatrix}
        1 & 0 \\
        0 & -1 \\
    \end{pmatrix}.
\end{align}
Here we assumed that the input beam is a purely s-polarized beam.
From Eqs. (\ref{eq:aftersample}) and (\ref{eq:output}), the electric field after transmitting both the birefringent sample and the HWP with $\theta_{\mathrm{H}}=0$ can be written as
\begin{align}
    \ket{E_{\mathrm{out}}(0)} &= E_{\mathrm{in}}\mathrm{e}^{i\alpha_+}
    \begin{pmatrix}
        \cos\alpha_- + i\sin\alpha_-\cos2\theta \\
        -i\sin\alpha_-\sin2\theta \\
    \end{pmatrix}.
    \label{eq:output_field}
\end{align}
The beam power for each polarization can be calculated as
\begin{align}
    |E_{\mathrm{in}}|^2
    \begin{pmatrix}
        \cos^2\alpha_- + \sin^2\alpha_-\cos^22\theta \\
        \sin^2\alpha_-\sin^22\theta \\
    \end{pmatrix}.
\end{align}
We detect the beam power of each polarization by using a polarized beam splitter (PBS) and photodiodes (PDs)
We define the normalized output beam power with the HWP angle of $\theta_{\mathrm{H}}$ as
\begin{align}
    V(\theta_{\mathrm{H}}) &\coloneqq 
    \begin{pmatrix}
        V_{\mathrm{s}}(\theta_{\mathrm{H}}) \\
        V_{\mathrm{p}}(\theta_{\mathrm{H}}) \\
    \end{pmatrix}.
\end{align}
When $\theta_{\mathrm{H}}=0$, the normalized output beam power becomes
\begin{align}
V(0) = 
    \begin{pmatrix}
        V_{\mathrm{s}}(0) \\
        V_{\mathrm{p}}(0) \\
    \end{pmatrix}
    &=
    \begin{pmatrix}
        \cos^2\alpha_- + \sin^2\alpha_-\cos^22\theta \\
        \sin^2\alpha_-\sin^22\theta \\
    \end{pmatrix}.
    \label{eq:BirefLoss}
\end{align}
Note that $\sin^2\alpha_-\sin^22\theta$ corresponds to the optical loss induced by the birefringence. 

Then, we rotate the HWP by $\pi/8$.
The Jones matrix of the HWP becomes
\begin{align}
    \hat{M}_{\mathrm{HWP}}(\pi/8) &= \frac{1}{\sqrt{2}}
    \begin{pmatrix}
        1 & 1 \\
        1 & -1 \\
    \end{pmatrix}.
\end{align}
One can compute the fields at PDs in the same way, and arrive at
\begin{align}
    \begin{pmatrix}
        V_{\mathrm{s}}(\pi/8) \\
        V_{\mathrm{p}}(\pi/8) \\
    \end{pmatrix}
    &= \frac{1}{2}
    \begin{pmatrix}
        \cos^2\alpha_- + \sin^2\alpha_-(\cos2\theta+\sin2\theta)^2 \\
        \cos^2\alpha_- + \sin^2\alpha_-(\cos2\theta-\sin2\theta)^2 \\
    \end{pmatrix}.
    \label{eq:normalized_pi8}
\end{align}
From Eq. (\ref{eq:normalized_pi8}), this can be rewritten as
\begin{align}
    V_{\mathrm{s}}(\pi/8) - V_{\mathrm{p}}(\pi/8) &= 2\sin^2\alpha_-\sin2\theta\cos2\theta.
\end{align}
By dividing $V_{\mathrm{p}}(0)$ by $V_{\mathrm{s}}(\pi/8) - V_{\mathrm{p}}(\pi/8)$, one can get
\begin{align}
    \frac{V_{\mathrm{p}}(0)}{V_{\mathrm{s}}(\pi/8) - V_{\mathrm{p}}(\pi/8)} &= \frac{1}{2}\tan2\theta.
\end{align}
Therefore, the orientation of the birefringence can be obtained from
\begin{align}
    \theta &= \frac{1}{2}\arctan\left(\frac{V_{\mathrm{p}}(0)}{V_{\mathrm{s}}(\pi/8) - V_{\mathrm{p}}(\pi/8)}\right).
    \label{eq:theta}
\end{align}
From $V_{\mathrm{p}}(0)$ and Eq. (\ref{eq:theta}), one can obtain the absolute value of $\alpha_-$ as
\begin{align}
    |\alpha_-| &= \arcsin\left(\frac{\sqrt{V_{\mathrm{p}}(0)}}{|\sin2\theta|}\right).
\end{align}
Similarly, one can extract the birefringence, $\theta$ and $|\alpha_-|$, when the input beam is p-polarization.

\section{Experimental setup}

\subsection{Configuration}

\begin{figure}
\includegraphics[width=8.6cm]{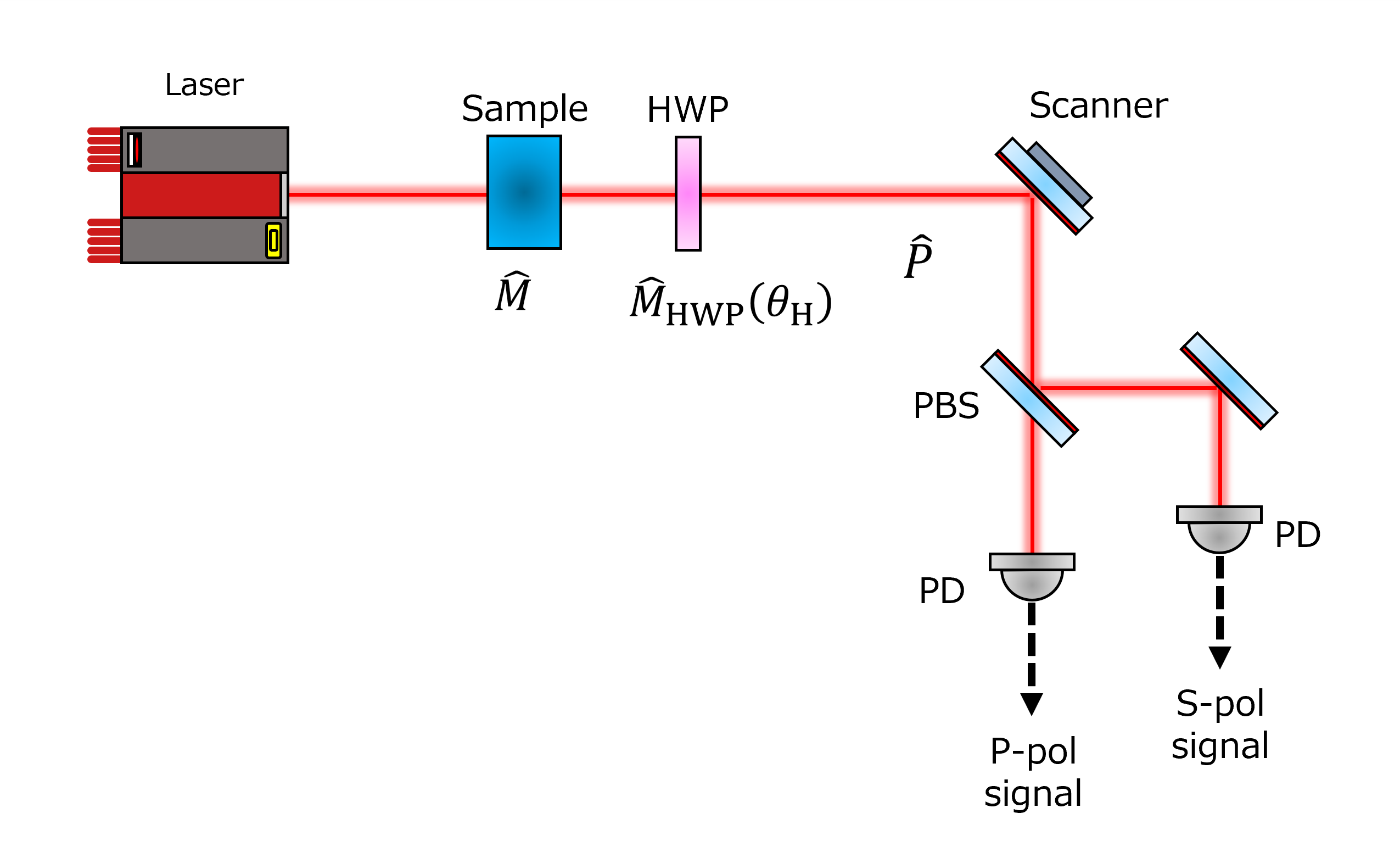}
\caption{\label{fig:PPC}
Simplified schematic of the experimental setup.
The input beam is conditioned to be s-polarized beam using polarizers (not shown in this figure).
After transmitting the HWP, the beam is split into s- and p-polarized beam using a PBS, and each beam power is detected by a PD.
}
\end{figure}

Fig. \ref{fig:PPC} shows our experimental setup.
A $1064\unit{nm}$ wavelength laser is employed here.
The input beam is an s-polarized Gaussian beam with a polarization extinction ratio of less than $0.1\%$.
The HWP is mounted on a rotational mount which has a vernier scale so that one can determine the HWP rotation angle precisely.
The output beam from the HWP is split into s- and p-polarized beams by a PBS.
Then, each beam is detected by a pin-hole PD which has an aperture of $55\unit{\mu m}$.
The distance between the scanner and each PD is set at $470\unit{mm}$.
The beam diameter on the sample and on each PD are about $2.6\unit{mm}$ and $1.2\unit{mm}$, respectively.

The beam is scanned by a scanner which is used as in the phase camera setup in Advanced Virgo. \cite{Schaaf2016, Agatsuma2019}
Similar to the Advanced Virgo setup, the following Archimedean spiral pattern is employed in our setup.
\begin{align}
    x(t) = R_{\mathrm{img}}\frac{t}{T_{\mathrm{img}}}\cos\left(2\pi f_{\mathrm{scan}}t\right), \\
    y(t) = R_{\mathrm{img}}\frac{t}{T_{\mathrm{img}}}\sin\left(2\pi f_{\mathrm{scan}}t\right).
\end{align}
$R_{\mathrm{img}}$, $T_{\mathrm{img}}$, and $f_{\mathrm{scan}}$ are the radius of image, the acquisition time for one image, and the spiral rotation frequency, respectively.
This scanning pattern enables a smooth trajectory, thus smooth mirror movement.
We scan over the beam diameter, i.e., $\sim1.2\unit{mm}$ in this setup, so that the birefringence distribution within the beam diameter can be characterized.

The output signal from each PD during scanning the beam is acquired by a digital system which is used in Advanced LIGO. \cite{Aasi2015}
Analog PD signals are converted to digital signals (counts) by 16 bit analog-to-digital converters (ADCs).
The signals are recorded as counts for each measurement.




\subsection{Propagator}

In reality, some optics are placed between the HWP and the PDs as shown in Fig. \ref{fig:PPC}.
For instance, a mirror on the scanner can have different transmissivities for s- and p-polarizations.
Such optical components affect the beam power at the PDs.
We define this effect as
\begin{align}
    V_{\mathrm{PD}} &= \hat{P}V(\theta_{\mathrm{H}}),
\end{align}
where
\begin{align}
    \hat{P} &\coloneqq
    \begin{pmatrix}
        P_{11} & P_{12} \\
        P_{21} & P_{22} \\
    \end{pmatrix},
\end{align}
denotes the propagation matrix from the HWP to the PDs.
In order to make a birefringence map, we need to convert $V_{\mathrm{PD}}$ to $V(\theta_{\mathrm{H}})$.
This can be done by multipling the inverse of $\hat{P}$ as
\begin{align}
    V(\theta_{\mathrm{H}}) &= \hat{P}^{-1}V_{\mathrm{PD}}.
\end{align}

The propagator, $\hat{P}$, can be measured with pure s- and p-polarzations without a sample.
When the output beam from the HWP has pure p-polarization
\begin{align}
    V_{\mathrm{PD}} &= \hat{P}
    \begin{pmatrix}
        1 \\
        0 \\
    \end{pmatrix},
    \notag \\
    &=
    \begin{pmatrix}
        P_{11} \\
        P_{21} \\
    \end{pmatrix}.
\end{align}
Similarly, when the input beam has pure s-polarization, one can obtain
\begin{align}
    V_{\mathrm{PD}}
    &=
    \begin{pmatrix}
        P_{12} \\
        P_{22} \\
    \end{pmatrix}.
\end{align}
From these two measurements, one can determine the propagation matrix $\hat{P}$.
Prior to the measurement with a sample, we characterized the propagator with this method.




\section{Results}

\begin{table}
\caption{\label{tab:results}
Summary of the results.
Each direction of the QWP was chosen at random.
The differential phase retardation of the QWP is fixed at $\alpha_-=\pi/4=0.785$.
Peak values are presented as measured values.
}
\begin{ruledtabular}
\begin{tabular}{cc|cc}
\multicolumn{2}{c|}{Estimated values} & \multicolumn{2}{c}{Measured values (Peak)}\\
$\theta\unit{[deg]}$ & $\alpha_-\unit{[rad]}$ & $\theta\unit{[deg]}$ & $\alpha_-\unit{[rad]}$ \\ \hline
$\approx19\unit{deg}$ & $\pi/4=0.785$ & $18.51$ & $0.789$ \\
 & & $19.31$ & $0.784$ \\
 & & $19.03$ & $0.788$ \\
 & & $18.81$ & $0.791$ \\
 & & $18.93$ & $0.794$ \\ \hline
$\approx34\unit{deg}$ & $\pi/4=0.785$ & $33.99$ & $0.771$ \\
 & & $33.99$ & $0.770$ \\
 & & $33.97$ & $0.770$ \\
 & & $33.91$ & $0.770$ \\
 & & $33.80$ & $0.771$ \\ \hline
$\approx-12\unit{deg}$ & $\pi/4=0.785$ & $-12.05$ & $0.791$ \\
 & & $-12.21$ & $0.790$ \\
 & & $-12.38$ & $0.784$ \\
 & & $-11.97$ & $0.791$ \\
 & & $-12.36$ & $0.776$ \\ 
\end{tabular}
\end{ruledtabular}
\end{table}

To verify the validity and repeatability of the proposed method, birefringence map of a quarter-wave plate (QWP) was measured as it has a fixed value of $|\alpha_-|=\pi/4$.
A zero-order QWP (WPQ10M-1064) from THORLABS is used as a sample in our setup. \cite{QWP}
Five sets of measurements were taken for each of the three different orientations.

\begin{figure*}[htbp]
\includegraphics[width=17.8cm]{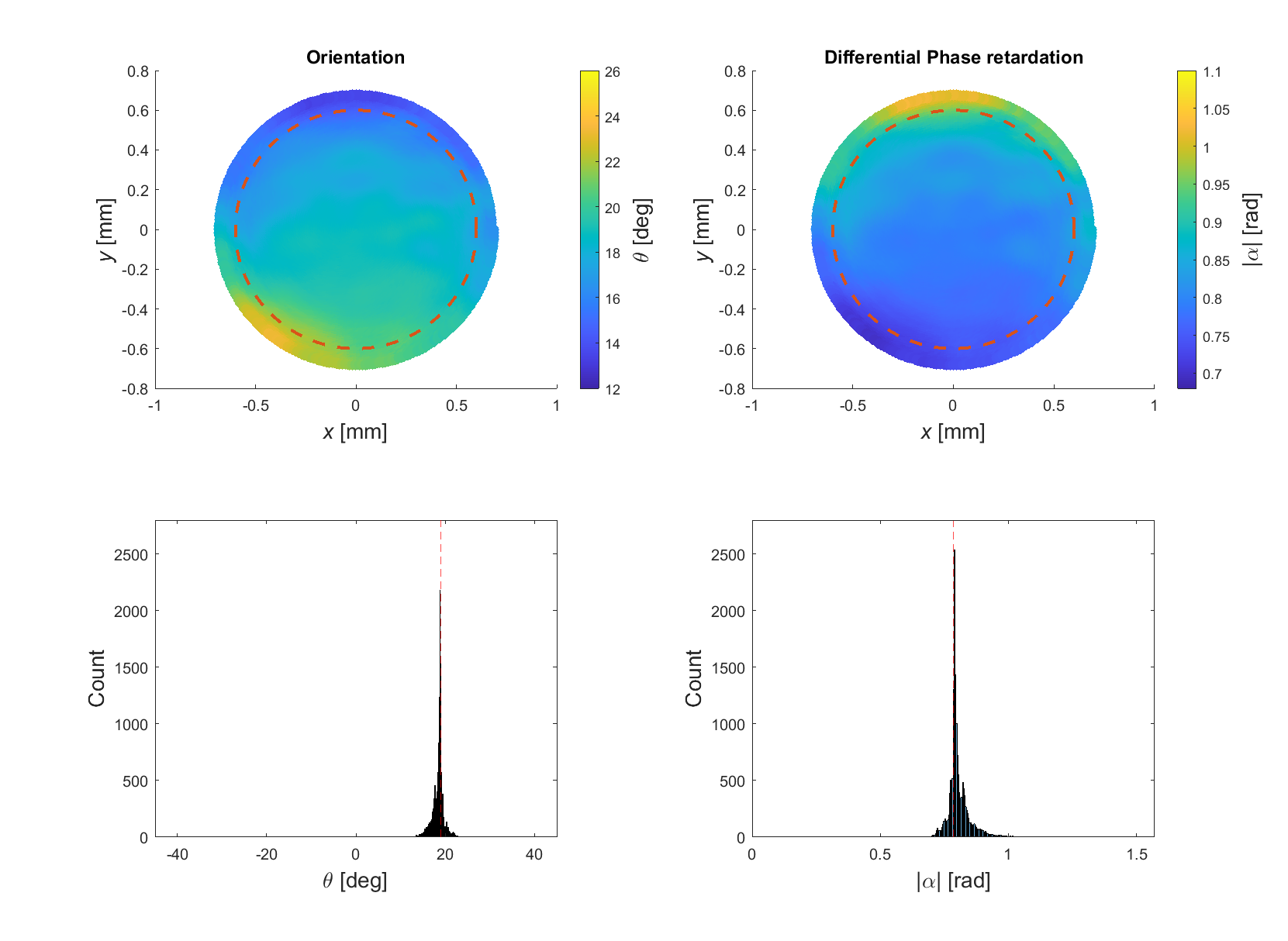}
\caption{\label{fig:QWPresutls}
Typical birefringence maps (upper) and their histogram (bottom) with the QWP.
Red dashed circles in the birefringence maps correspond to the beam diameter at the PDs ($1.2\unit{mm}$).
Red dashed lines in the histograms are the estimated values $19\unit{deg}$ and $\pi/4$, respectively.
The histograms indicate that the measured values are consistent with the estimated values.
}
\end{figure*}

Table \ref{tab:results} summarizes the obtained results, and Fig. \ref{fig:QWPresutls} shows the typical result of the birefringence distributions and their histograms.
In Table \ref{tab:results}, we describe the values of $\theta$ and $\alpha_-$ which give the peak values in their histograms.
Both the orientation and the differential phase retardation can be measured within $\sim3\%$ variations.
Within the beam diameter where the beam is most sensitive to disturbance such as birefringence, the measured orientation and differential phase retardation show reasonable homogeneity ($\sim3\%$).

Measured results indicate that the proposed characterization method has reasonable repeatability regardless of the axis orientation.
The sapphire substrate used in KAGRA showed the one-way differential phase change of $\alpha_-\approx0.6\unit{rad}$. \cite{Wang2024}
Therefore, our setup has sufficient sensitivity and repeatability for in-situ birefringence analysis in current KAGRA sapphire mirrors.
For the case of next-generation GWDs such as Cosmic Explorer, a $38\unit{cm}$-thick fused silica substrate is planned to be employed with a $1064\unit{nm}$ wavelength main laser.
Assuming the birefringence of such fused silica is $\Delta n=10^{-7}$, the differential phase is $\alpha_-\approx0.22\unit{rad}$ per round trip, which is in the same order of the measured sample in our setup ($\alpha_-\approx0.78\unit{rad}$).
Therefore, the proposed scheme can be useful even for Cosmic Explorer.
Further measurements with a fused silica substrate which has a known amount of birefringence will pave the path toward application for next-generation GWDs.

\subsection{Discussions on the error}

We discuss possible error sources which might limit the performance of the proposed scheme.

\subsubsection{Input beam polarization extinction ratio}

The input beam extinction ratio needs to be kept as low as possible as it can couple to $V_{\mathrm{h}}(0)$. 
As long as the input beam extinction ratio is much smaller than $\sin^2\alpha_-\sin^22\theta$, the impacts on the birefringence measurement can be ignored.
For instance, $V_{\mathrm{h}}(0)$ is typically in order of $10^{-1}$, therefore, as long as the input beam extinction ratio is kept below $10^{-3}$, it will not largely affect the birefringence characterization.
This has been achieved in the GWDs.
For instance, in Advanced LIGO, the input beam polarization extinction ratio is $10^{-5}$. \cite{Mueller2016}

\subsubsection{HWP misalignment}
When the HWP axis has a small misalignment, $\delta\theta_{\mathrm{H}}(\ll1)$, it introduces a systematic error in the birefringence map.
This effect can be estimated as follows.
$V(0)$ in Eq. (\ref{eq:BirefLoss}) can be replaced by $V(\delta\theta_{\mathrm{H}})$.
Assuming that $\delta\theta_{\mathrm{H}}\ll1$, we can approximate the Jones matrix of the HWP as
\begin{align}
    \hat{M}_{\mathrm{HWP}}(\delta\theta_{\mathrm{H}}) &=
    \begin{pmatrix}
        1-2\sin^2(\delta\theta_{\mathrm{H}}) & \sin(2\delta\theta_{\mathrm{H}}) \\
        \sin(2\delta\theta_{\mathrm{H}}) & -1+2\sin^2(\delta\theta_{\mathrm{H}}) \\
    \end{pmatrix}, \notag \\
    &\approx
    \begin{pmatrix}
        1 & 2\delta\theta_{\mathrm{H}} \\
        2\delta\theta_{\mathrm{H}} & -1 \\
    \end{pmatrix}.
\end{align}
Here we ignore the second-order terms with respect to $\delta\theta_{\mathrm{H}}$.
From Eq. (\ref{eq:output_field}), the output field with small HWP misalignment can be expressed as
\begin{widetext}
\begin{align}
    \ket{E_{\mathrm{out}}(\delta\theta_{\mathrm{H}})} &= E_{\mathrm{in}}\mathrm{e}^{i\alpha_+}
    \begin{pmatrix}
        \cos\alpha_- + i\sin\alpha_-\cos2\theta + 2i\delta\theta_{\mathrm{H}}\sin2\theta\ \\
        -i\sin\alpha_-\sin2\theta + 2\delta\theta_{\mathrm{H}}(\cos\alpha_--i\sin\alpha_-\cos2\theta) \\
    \end{pmatrix}.
\end{align}
\end{widetext}
Then, the normalized beam power can be computed as
\begin{align}
    V(\delta\theta_{\mathrm{H}}) & \approx
    \begin{pmatrix}
        \cos^2\alpha_- + \sin^2\alpha_-\cos^22\theta \\
        \sin^2\alpha_-\sin^22\theta + 4\delta\theta_{\mathrm{H}}\sin^2\alpha_-\sin2\theta\cos2\theta \\
    \end{pmatrix}.
\end{align}
As long as $4\delta\theta_{\mathrm{H}}\sin^2\alpha_-\sin2\theta\cos2\theta$ is much smaller than $\sin^2\alpha_-\sin^22\theta$, the impact of the HWP misalignment can be negligible.
This requirement can be rewritten as
\begin{align}
    \frac{4\delta\theta_{\mathrm{H}}\sin^2\alpha_-\sin2\theta\cos2\theta}{\sin^2\alpha_-\sin^22\theta} = \frac{4\delta\theta_{\mathrm{H}}}{\tan2\theta} \ll1.
\end{align}
Typically, $\tan2\theta$ is in the order of unity.
Therefore, as long as $\delta\theta_{\mathrm{H}}$ should be kept below a few mrad, i.e., $\lesssim0.2\unit{deg}$, the HWP misalignment will not significantly affect the results.
This requirement can be achieved even with a motorized rotation mount which has $0.1\unit{mrad}$ minimum incremental motion. \cite{Newport}



\section{Implementation details}

\subsection{A possible characterization setup in a GWD}

\begin{figure}[htbp]
\centering
\includegraphics[width=8.6cm]{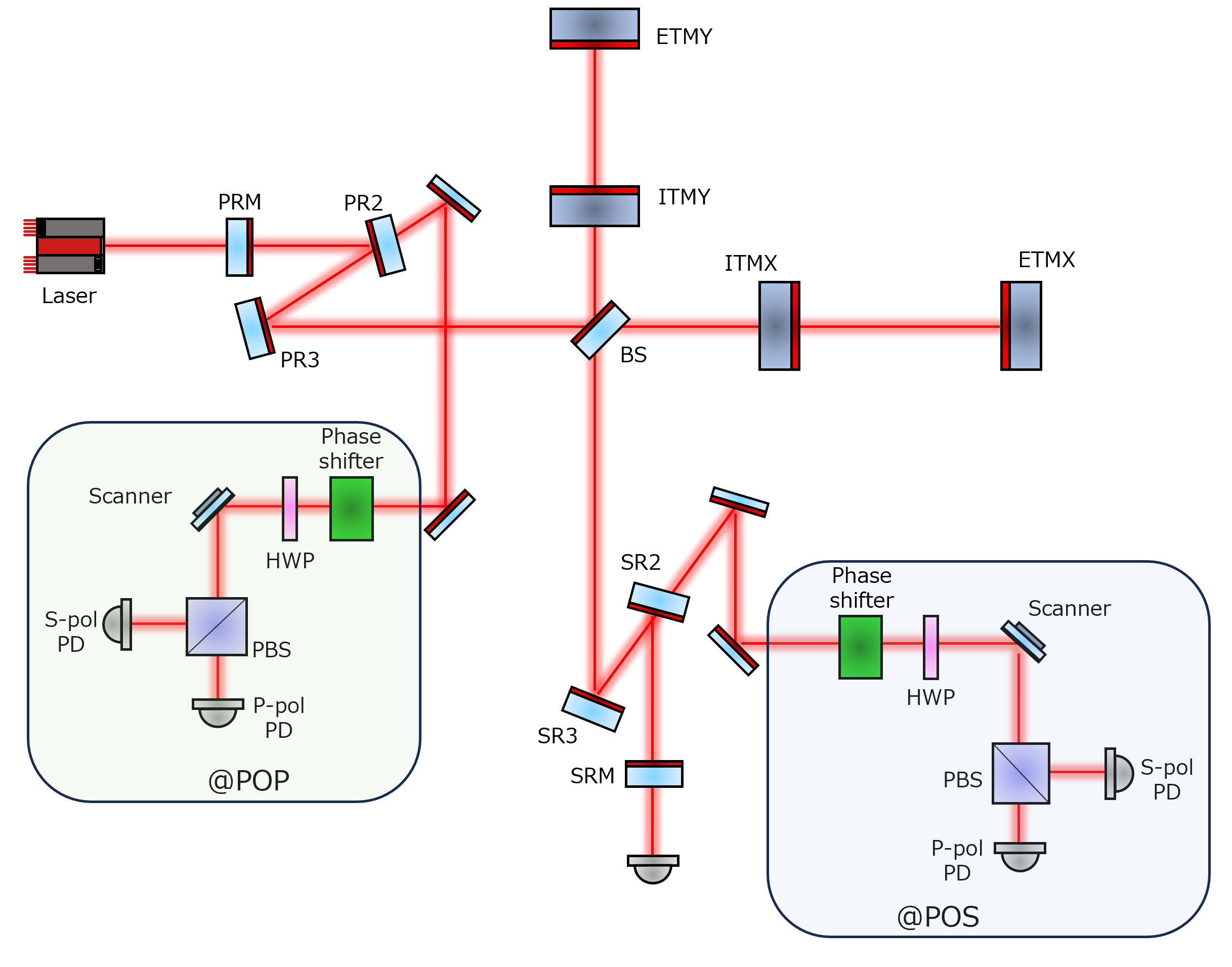}
\caption{Simplified schematic of the possible birefringence characterization setup in a gravitational wave detector.
For instance, KAGRA has two optical tables, POP and POS, where the laser beams from the power recycling and signal recycling cavities are picked off.
When the proposed polarimetry setup is installed on each table, then can be used as a diagnostic tool to investigate the birefringence property of the GWD.
}
\label{fig:setupGWD}
\end{figure}

The proposed scheme will enable us to characterize the birefringence in GWDs.
Fig. \ref{fig:setupGWD} shows a possible birefringence characterization setup in a GWD.
For instance, KAGRA has two tables, called POP and POS, which pick off the main beam from the power recycling and signal recycling cavities.
These picked off beams are affected by the test mass birefringence.
By installing the proposed polarimetry setup on each table and analyzing the picked off beams, we will be able to investigate the birefringence in each recycling cavity without introducing an auxiliary light source.
Since this method utilize the main beam of the interferometer, it can be applied to the Einstein Telescope that plans to use a $1550\unit{nm}$ wavelength laser. \cite{Punturo2010}
Note that a phase shifter is installed in front of the HWP on each table whose role is described in the following subsection.

This setup can also be used to measure the distribution of the birefringent optical loss.
When the beam passes through a birefringent material, a certain amount of beam power is scattered into orthogonal polarization as show in Eq. (\ref{eq:BirefLoss}), which can be regarded as an optical loss.
If the birefringent loss has asymmetry between two arms, it can degrade the detector sensitivity through the contrast defect.
Furthermore, such birefringent loss can degrade the performance of the squeezing which reduces the quantum noise. \cite{Krüger2016, Michimura2024}
Therefore, the polarimetry setup proposed in this manuscript serve as a diagnostic tool to assess the birefringent loss in GWDs.

\subsection{Compensation of optics}

Even though the birefringence mapping scheme is expected to enable the \textit{in-situ} characterization of birefringence distribution and birefringent optical loss, it requires further developments for implementation.

\subsubsection{Transmissivity imbalance}
In our tabletop setup, the HWP is placed immediately after the sample.
However, in a GWD, several optical components are placed between the ITM and the HWP.
Such optics, for instance the beamsplitter (BS), introduces imbalanced transmissivity between s- and p-polarized beams.
This results in an imbalance in the beam power at each PD.
However, this effect can be addressed by a post process as long as the properties of optical components are known in advance.

We define the propagation matrix which introduces optical attenuation as
\begin{align}
    \hat{M}_{\mathrm{prop}} &=
    \begin{pmatrix}
        \mathfrak{t}_{\mathrm{s}} & 0 \\
        0 & \mathfrak{t}_{\mathrm{p}}
    \end{pmatrix},
\end{align}
where $\mathfrak{t}_{\mathrm{s}}$ and $\mathfrak{t}_{\mathrm{p}}$ are the amplitude transmissivities from the sample to the HWP.
Here we assume that the birefringence in these optics can be negligible.
Taking this propagation into account, the output fields are modified as
\begin{align}
    \ket{E_{\mathrm{out}}(\theta_{\mathrm{H}})} &= \hat{M}_{\mathrm{HWP}}(\theta_{\mathrm{H}})\hat{M}_{\mathrm{prop}}\hat{M}\ket{E_{\mathrm{in}}}.
\end{align}
The normalized beam power can be modified as
\begin{align}
    V'(0) &= 
    \begin{pmatrix}
        \mathfrak{t}_{\mathrm{s}}^2\cos^2\alpha_- + \sin^2\alpha_-\cos^22\theta \\
        \mathfrak{t}_{\mathrm{p}}^2\sin^2\alpha_-\sin^22\theta \\
    \end{pmatrix}, \\
    V'(\pi/8) &= \frac{1}{2}
    \begin{pmatrix}
        \mathfrak{t}_{\mathrm{s}}^2\cos^2\alpha_- + \sin^2\alpha_-(\mathfrak{t}_{\mathrm{s}}\cos2\theta+\mathfrak{t}_{\mathrm{p}}\sin2\theta)^2 \\
        \mathfrak{t}_{\mathrm{s}}^2\cos^2\alpha_- + \sin^2\alpha_-(\mathfrak{t}_{\mathrm{s}}\cos2\theta-\mathfrak{t}_{\mathrm{p}}\sin2\theta)^2 \\
    \end{pmatrix}.
\end{align}
These lead to changes in results as follows:
\begin{align}
    \frac{V'_{\mathrm{p}}(0)}{V'_{\mathrm{s}}(\pi/8) - V'_{\mathrm{p}}(\pi/8)} &= \frac{\mathfrak{t}_{\mathrm{p}}}{2\mathfrak{t}_{\mathrm{s}}}\tan2\theta.
\end{align}
Thus,
\begin{align}
    \theta &= \frac{1}{2}\arctan\left(2\frac{\mathfrak{t}_{\mathrm{s}}}{\mathfrak{t}_{\mathrm{p}}}\frac{V'_{\mathrm{p}}(0)}{V'_{\mathrm{s}}(\pi/8) - V'_{\mathrm{p}}(\pi/8)}\right).
\end{align}
As long as the optical properties are properly characterized, we can compensate the transmissivity imbalance in post processing.
The absolute value of differential phase retardation, $|\alpha_-|$, can also be derived in the same way.
\begin{align}
    |\alpha_-| &= \arcsin\left(\frac{\sqrt{V'_{\mathrm{p}}(0)}}{|\mathfrak{t}_{\mathrm{p}}\sin2\theta|}\right).
\end{align}

\subsubsection{Relative phase delay}

The BS introduces not only transmissivity imbalance, but also a relative phase delay between s- and p-polarized beams. \cite{Michimura_note}
The Jones matrix of such relative phase delay can be expressed as
\begin{align}
    \hat{M}_{\mathrm{delay}}
    &=
    \begin{pmatrix}
        1 & 0 \\
        0 & \mathrm{e}^{i\phi} \\
    \end{pmatrix},
\end{align}
where $\phi$ denotes the relative phase delay.
Then, Eq. (\ref{eq:output}) can be rewritten as
\begin{align}
    \ket{\tilde{E}_{\mathrm{out}}(\theta_{\mathrm{H}})} &= \hat{R}(\theta_{\mathrm{H}})\hat{M}_{\mathrm{delay}}\hat{M}{\boldsymbol{E}}_{\mathrm{in}}.
    \label{eq:PhaseDelay}
\end{align}
Therefore, the results can be affected by this phase delay.
When $\theta_{\mathrm{H}}=0$, the normalized beam powers do not change.
However, when we rotate the HWP such that $\theta_{\mathrm{H}}=\pi/8$, the above equation becomes
\begin{align}
    \ket{\tilde{E}_{\mathrm{out}}(\theta_{\mathrm{H}})} &= \hat{R}(\pi/8)\hat{M}_{\mathrm{delay}}\hat{M}{\boldsymbol{E}}_{\mathrm{in}}.
\end{align}
By doing some math, the normalized beam power can be calculated as
\begin{widetext}
\begin{align}
    \begin{pmatrix}
        \tilde{V}_{\mathrm{s}}(\pi/8) \\
        \tilde{V}_{\mathrm{p}}(\pi/8) \\
    \end{pmatrix}
    &= \frac{1}{2}
    \begin{pmatrix}
        1-2\cos\alpha_-\sin\alpha_-\sin2\theta\sin\phi + 2\sin^2\alpha_-\sin2\theta\cos2\theta\cos\phi \\
        1+2\cos\alpha_-\sin\alpha_-\sin2\theta\sin\phi - 2\sin^2\alpha_-\sin2\theta\cos2\theta\cos\phi \\
    \end{pmatrix}.
    \label{eq:normalized_pi8withdelay}
\end{align}
\end{widetext}
\newpage
When the beam experiences some relative phase delay between the sample and the HWP, the differential normalized beam power between s- and p-polarizations becomes
\begin{align}
    \tilde{V}_{\mathrm{s}}(\pi/8) - \tilde{V}_{\mathrm{p}}(\pi/8) &= 2\sin^2\alpha_-\sin2\theta\cos2\theta\cos\phi \notag \\ 
    &\quad\quad - 2\sin\alpha_-\cos\alpha_-\sin2\theta\sin\phi.
\end{align}
Even if the phase delay of $\phi$ is known, one cannot derive the birefringence, $\alpha_-$ and $\theta$, as $\tilde{V}_{\mathrm{s}} - \tilde{V}_{\mathrm{p}}$ is a nonlinear function.



Although the relative phase delay in the optics prevent us from birefringence characterization, this can be addressed by inserting a variable phase shifter such as an electro-optic modulator (EOM) in front of the HWP. \cite{Kaneshiro2016}
The Jones matrix of a phase shifter can be expressed as
\begin{align}
    \hat{M}_{\mathrm{PS}}
    &=
    \begin{pmatrix}
        1 & 0 \\
        0 & \mathrm{e}^{i\varepsilon} \\
    \end{pmatrix},
\end{align}
where $\varepsilon$ is the relative phase between s- and p-polarizations.
Here, we ignore the transmissivity of the phase shifter for each polarization as it can be compensated as shown in above.
The phase retardation, $\varepsilon$, can be easily controlled by changing the applied voltage to the retarder.
Once we know the phase delay, $\phi$, we can set $\varepsilon=-\phi$ so that we can eliminate its impact.
Then, the proposed birefringence characterization method can be applied.
While the EOM can be used as the phase shifter, if there is significant non-uniform birefringence in the EOM, the relative phase delay may not be fully compensated.
In addition, the phase shifter may need to be calibrated occasionally since the phase retardation, $\varepsilon$, may change over time.

We are planning to perform further study and verification of the compensation schemes for both transmissivity imbalance and relative phase delay.
Specifically, the impacts of the EOM as the phase shifter will be evaluated in the future.

\section{Conclusion}

A novel scheme for \textit{in-situ} birefringence characterization of a gravitational wave detector has been proposed and demonstrated with a tabletop experiment.
We have shown the working principle of the birefringence mapping scheme which is compatible with a scanning pin-hole phase camera used in the gravitational wave detector.
Therefore, the proposed polarimetry birefringence mapping method is able to characterize the \textit{in-situ} birefringence distribution and birefringent optical loss in a gravitational interferometer.

In order to implement this system in a GWD, phase-delay and transmissivity imbalance introduced by the optics need to be assessed and compensated.
We are currently planning to perform an experimental verification of the compensation scheme and demonstration for implementation in GWDs.

\begin{acknowledgments}

The authors thank Daniel Brown, Evan Hall, Kevin Kuns, and the LIGO Scientific Collaboration Advanced Interferometer Configuration working group for useful discussions and feedback. 
The authors thank Alberto Vecchio, Stephen Webster, and Martin van Beuzekom for their contributions to the instrumental acquisition. 
KK acknowledges Kazuhiro Agatsuma, and Hiroaki Yamamoto for their inputs to the project.
This work was supported by the Royal Society's Research Grant, RGS\textbackslash R2\textbackslash212142.
TP gratefully acknowledges the UKRI STFC Student Grant (ST/W507374/1). 
The authors appreciate Cardiff University's support in making this research possible.
This paper has LIGO Document number LIGO-P2500205.
\end{acknowledgments}

\section*{Author Declarations}

\subsection*{Conflict of Interest}
The authors have no conflicts to disclose.


\section*{Data Availability Statement}

The data that support the findings of this study are available from the corresponding author upon reasonable request.



\nocite{*}
\bibliography{aipsamp}

\end{document}